\documentclass[a4paper]{llncs}

\usepackage{amssymb}
\usepackage{graphicx}
\usepackage{xcolor} 
\usepackage[normalem]{ulem} 
\usepackage{wrapfig} 
\usepackage{enumerate}
\usepackage[shortlabels]{enumitem}
\usepackage{todonotes}
\usepackage{pifont}
\usepackage{enumitem}
\usepackage{tikz}
\usetikzlibrary{positioning,fit}
\usepackage[utf8]{inputenc}
\usepackage[T1]{fontenc}
\usepackage{tgtermes}
\usepackage[scaled=0.93]{beramono}
\usepackage[normalem]{ulem}
\usepackage{cite}
\usepackage{float}
\usepackage{scalefnt}
\usepackage{setspace}
\usepackage{caption}
\usepackage{subcaption}
\usepackage{booktabs}

\usepackage{url}
\urldef{\mailsa}\path|{johan.linaker,bjorn.regnell}@cs.lth.se|
\newcommand{\keywords}[1]{\par\addvspace\baselineskip
\noindent\keywordname\enspace\ignorespaces#1}

\begin{document}

\mainmatter  

\title{A Contribution Management Framework for Firms\\
Engaged in Open Source Software Ecosystems\\
- A Research Preview}

\titlerunning{A Contribution Management Framework - A Research Preview}

%
%
\author{Johan Lin{\aa}ker and Bj{\"o}rn Regnell}
\authorrunning{Johan Lin{\aa}ker and Bj{\"o}rn Regnell}

\institute{Lund University, Lund, Sweden\\
\mailsa}

%
%

\toctitle{Lecture Notes in Computer Science}
\tocauthor{Authors' Instructions}
\maketitle

\vspace{-0.5cm}
\begin{abstract}
[\textbf{Context and motivation}]
Contribution Management helps firms engaged in Open Source Software (OSS) ecosystems to create contribution strategies which motivate what they should contribute and when, but also what they should focus their resources on and to what extent. The motivation for developing tailored contribution strategies is to maximize return on investment and sustain the influence needed in the ecosystem.
[\textbf{Question/Problem}] 
We aim to develop a framework to help firms understand their current situation and create a starting point to develop an effective contribution management process. 
[\textbf{Principal ideas/results}] 
Through a design science approach, a prototype framework is created based on literature and validated iteratively with expert opinions through interviews.
[\textbf{Contribution}] 
In this research preview, we present our initial results after our first design cycle and consultation with one experienced OSS manager at a large OSS oriented software-intensive firm. The initial validation highlights importance of stakeholder identification and analysis, as well as the general need for contribution management and alignment with internal product planning. This encourages future work to develop the framework further using expert and case validation.

\end{abstract}

\section{Introduction}
\vspace{-0.3cm}


Requirements Engineering (RE) concerns capturing the needs of the customer and translating these into a product that satisfies the elicited needs~\cite{aurum2005requirements}. For software-intensive firms, RE can therefore be considered as a pivotal part in the product planning and spans over different time horizons and abstractions, from requirements management, to release-planning, roadmapping and portfolio management~\cite{fricker2012software}. Firms operating in an Open Source Software (OSS) ecosystem have to consider participation in two such RE instances; one that regards the internal product planning facilitated by themselves, and one that regards the external product planning of the OSS project facilitated by the OSS ecosystem. In the latter, to impose their own agendas, firms have to collaborate and compete with other actors in the ecosystem that all have a stake in the OSS project that underpins the ecosystem~\cite{jansen2009business}. 

To gain the influence needed in order to impose their agenda, align their internal RE with the ecosystem's RE and to maximize Return On Investment (ROI), firms need consider how to participate in the OSS ecosystem in terms of what they contribute and when (cf. requirements scoping~\cite{wnuk2012can}), but also what they should focus their resources on and to what extent. We choose to label this process as \emph{contribution management} and guidelines that come as output \emph{contribution strategies}~\cite{wnuk2012can}. To create these strategies, we believe that firms need to understand how they draw value~\cite{Aurum2007} from the OSS project and their ecosystems~\cite{jansen2009business}, and identify the related business requirements~\cite{wiegers2013software}. Further, firms need to understand the relation between the OSS projects, their ecosystems, and the firms' internal product planning~\cite{fricker2012software}, and as a consequence how important is it to be able to influence the RE in the OSS ecosystems~\cite{linaaker2016firms}. These factors need to align with the reasons for why firms make contributions and dedicate resources to the OSS project and its ecosystem, i.e., the foundations for the contribution management process.

We aim to develop a contribution management framework to help firms understand their current situation and create a starting point that can help them construct guidelines for what they should contribute to the OSS ecosystems and when, i.e., contribution strategies~\cite{wnuk2012can}. We apply a design science approach~\cite{wieringa2014design} by first building on literature~\cite{munir2015open}, and then consult with experts for opinions in an iterative fashion. In this research preview we present our initial results after our first design cycle and consultation with an experienced OSS manager at a large OSS oriented software-intensive firm.

\vspace{-0.3cm}
\section{Research Methodology}
\vspace{-0.3cm}

We consider the problem context of aligning the contribution management with a firm's internal product planning~\cite{fricker2012software} and business requirements~\cite{wiegers2013software} as a design problem. We adopt a design science approach~\cite{wieringa2014design} and define our design problem to:

\begin{itemize}[nosep]\footnotesize
    \item \textit{Improve alignment between a firm’s contribution management towards OSS projects with the firm’s internal product planning and business requirements,
    by designing a framework
    that can help the firm to create guidelines for what should be contributed and when,
    in order for its developers to better decide what to contribute and how to prioritize their work. }
\end{itemize}

The treatment addressing the stated design problem includes the framework (i.e., the artifact~\cite{wieringa2014design}) as well as the interaction between it and the problem context, which in our case is constituted by firms engaged in an OSS ecosystem.


In our study, we identify and explain the problem based on literature~\cite{munir2015open} and develop a prototype of an artifact along with a potential interaction set-up. As a validation model, we will use expert opinions where the prototype and interaction set-up is simulated through interviews. Based on the output of each interview, the treatment is refined and again validated in a new cycle. The interviews are semi-structured with introductory questions that covers current involvement in OSS ecosystems, contribution practices, and how internal RE functions relative the ecosystems'. In the second part, the framework is presented for the interviewee with an explanation and open discussion on its structure and content. In the third part, the interviewee and interviewer walk through the framework for an OSS ecosystem of the interviewee's choice. The interview ends with a discussion of usage scenarios, potential improvements and changes of the framework. The interviews are audio-recorded and transcribed.

In this research preview, the results from the first design cycle is presented where our initial treatment design was validated with an OSS manager of a large OSS oriented software-intensive firm. Based on the output from the first interview, some factors in the original prototype was reordered and made clearer. E.g., the engagement and revealing strategies, originally elicited from Dahlander et al.~\cite{dahlander2008firms} was made more explicit, while the ecosystem stakeholder analysis and identification was abstracted as a general input to all levels. For the framework, see Fig.~\ref{fig:framework_picture}.

\vspace{-0.3cm}
\section{Structure of the Framework}
\vspace{-0.3cm}

\begin{figure*}[t!]
\vspace{-0.5cm}
\centering
\includegraphics[scale=0.45]{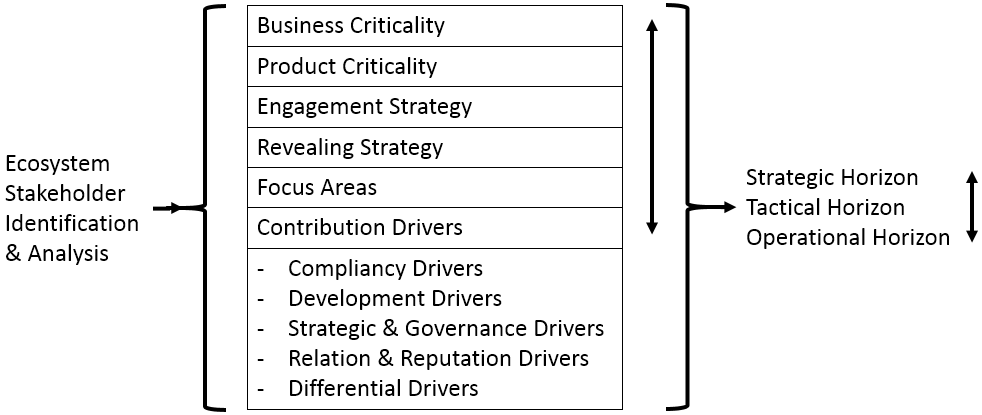}
\caption{Overview of the first iteration of the proposed contribution management framework.}
\label{fig:framework_picture}
\vspace{-0.5cm}
\end{figure*}

In this section we present the structure of the framework. As illustrated in Fig.~\ref{fig:framework_picture}, it consists of six levels: Business Criticality, Product Criticality, Engagement Strategy, Revealing Strategy, Focus Areas, and Contribution Drivers. These levels are used to frame and explain how the firm engages or should engage with a specific OSS ecosystem. Business and Product Criticality represents the role and importance of the OSS project and its ecosystem in relation to the firm's internal business requirements and product planning. Engagement and Revealing Strategy represents the way in how the firm interacts and contributes to OSS project and its ecosystem. Focus areas is used to separate between parts of an OSS project which are valued differently in terms of previous framework levels. Contributions Drivers are what motivate what, when and to whom a software artifact should be revealed, or where resources should be invested and to what extent, i.e., contribution management. The framework presents a list of possible drivers, but not all may be relevant or even listed. These should be identified and be in alignment with previous framework levels. As input to all framework levels, the firm should perform stakeholder identification and analysis on the OSS ecosystem. The output from the framework (i.e., how an OSS project is viewed in terms of Business and Product Criticality, what strategies are used, and what drivers that are relevant for which focus areas) may differ with time and should be in alignment with the firm's internal product planning, why it can be divided into different time horizons (e.g., Strategic, Tactical, and Operational). Below we present each part of the framework in more detail.

\noindent\textbf{Time Horizons:} To capture the short to long-range views, three horizons are defined: (1) the strategic horizon which looks beyond one year, (2) the tactical horizon which looks up to one year ahead, and (3) the operational horizon which is the practice at the current point in time. The precise time intervals are to be adapted relative each firm's internal product planning~\cite{fricker2012software}. The important aspect is how the engagement with the ecosystem should be adapted as time and development progresses.

\noindent\textbf{Ecosystem Stakeholder Population:} What stakeholders that are present and what their agendas are may affect how a firm judges the business criticality of the OSS project, to what extent the firm should engage the ecosystem, and what they choose to reveal, and when~\cite{munir2015open}. E.g., The presence of competitors may affect what is to be considered differential and not. Some stakeholders may be unknown and indirect competitors pending on their agendas~\cite{van2009commodification}. Similarly, the presence of existing and potential partners may offer opportunities for closer collaborations, some of which are too specific or differential to share with the rest of the ecosystem. Knowing who are the most influential and what their interests are may hint how the OSS project's roadmap aligns with the firm’s, how easy it is to affect, but also who that should be influenced to create traction in a direction favorable for the firm~\cite{linaaker2016firms}. Additionally, it may provide an input to if the OSS ecosystem is worth engaging in the first place, and also to help monitor the general health of it~\cite{jansen2009business}.

\noindent\textbf{Business Criticality:} Refers to how the firm draws value~\cite{Aurum2007} from the OSS project and its ecosystem, and how the related business requirements~\cite{wiegers2013software} are defined. From a business model perspective, the business criticality of the OSS may be judged based on the rationale of how it helps the firm to create, deliver, and capture value~\cite{osterwalder2010business}. E.g.,~\cite{chesbrough2007open, west2003open}, as a direct part of the product offering through an open core or platform-extension model, as a basis for support, subscriptions and professional services, or as part of a duel-licensing model. However, it may also be the case that the value comes indirect when the OSS is used as an enabler for the firms’ product offerings, e.g., as a development component or as part in the infrastructure supporting the product. It may also a combination of such direct and indirect factors. E.g., in asymmetric business models, software is made OSS to instead capture value from additional products, services and data gathering that is managed through the OSS~\cite{visionmobile2014assymetric}. Even though considered a difficult process~\cite{Komssi2015}, firms must be able to determine the strategic importance of the OSS in regard to differentiation and added value~\cite{Aurum2007} in order to decide if and how much the firm should invest and interact with the ecosystem~\cite{wnuk2012can}. 

\noindent\textbf{Product Criticality:} Refers to how the firm uses the OSS project in their future plans and actions in regards to their product over a series of releases, i.e., how integrated the OSS project is with the product and how the internal product planning~\cite{fricker2012software} needs to align with that of the OSS project. This affects what requirements need to be present in both or separately, and therefore what should be contributed or not. Further, if the firm uses a product-line approach with an underlying platform from which it creates its products, there may be an interest to contribute back to the OSS project in order to enable reuse. If they focus on developing single products and reuse more opportunistic, there may be less of a long-term perspective so less may be contributed back. 

\noindent\textbf{Engagement strategy:} Pending on the business and product criticality of the OSS project, the firm may need to have an influence on the development going on in the ecosystem. By actively engaging and contributing back to the ecosystem firms can increase their level of influence. Dahlander \& Magnusson~\cite{dahlander2005relationships} describes three types of relationships in regards to activity and influence on the ecosystem. Firstly, symbiotic relationships imply giving back to the ecosystem and is associated with a high influence for the firm. Second, commensalistic relationships imply interacting with the ecosystem but to the required minimum, and is associated with a low influence for the firm. Finally, parasitic relationships imply no interaction or giving back to the ecosystem, and is related to no, or very limited influence. Dahlander \& Magnusson~\cite{dahlander2005relationships} highlights that these are to be considered as a continuum. 

\noindent\textbf{Revealing strategy:} Pending on the business and product criticality, and the level of engagement, different strategies may be enforced in regards of what to reveal. E.g., by selectively revealing, differentiating parts can be kept closed while commodity parts can be made open~\cite{henkel2006selective, van2009commodification}. Further, with licensing schemas (cf. Dual-licensing~\cite{chesbrough2007open}), parts can be opened fully but under such circumstances that competitors cannot exploit the OSS that may hurt the focal firm~\cite{west2003open}. Alternatively, everything may be disclosed under open and transparent conditions~\cite{chesbrough2007open}, or even closed for that matter. Different strategies may be applied to different parts of an OSS project, as well as combined. 

\noindent\textbf{Focus Area(s):} Areas or modules of strategic importance and/or extra value to the firm. For some OSS projects, it may necessary to consider different parts or sub projects separately in regards to this framework.

\noindent\textbf{Contribution Drivers:} Pending on previous levels in the framework, the firm can identify which drivers that motivate what should be contributed and shared with the OSS ecosystem, and when, but also what resources should be dedicated and to what extent. Those listed in table~\ref{tbl:drivers} are not to be considered exhaustive, nor all relevant per default. There may be further drivers which are specific for the focal firm and how it makes use of the OSS project and its ecosystem.

\begin{table}[t!]
\centering
\caption{Contribution Drivers for why to contribute to an OSS project and ecosystem.}
\label{tbl:drivers}
\begin{tabular}{p{12.1cm}}
    \toprule
    \footnotesize
\textbf{Compliancy Drivers} 
    \\ 
    \midrule
    \footnotesize
    Parts required for compliance with licenses, patents, standards, and law.        
    \\
    \midrule
    \footnotesize
\textbf{Development Drivers}                                              
    \\     
    \midrule 
    \footnotesize
    Parts that can ease future maintenance and avoid unnecessary internal patch-work 
    \\ 
    \footnotesize
    Parts that may allow for better synced release cycles
    \\ 
    \footnotesize
    Parts that may reduce integration costs
    \\ 
    \footnotesize
    Parts that allows for third party products and services
    \\ 
    \footnotesize
    Parts that would benefit from external development and testing due to lack of internal resources, or a wish for increased quality and innovation
    \\ 
    \footnotesize
    Parts necessary to maintain an absorptive and learning capacity
    \\ 
    \footnotesize
    Parts necessary to keep a low entrance barrier for new developers
    \\ 
    \midrule
    \footnotesize
\textbf{Strategic \& Governance Drivers:}
    \\ 
    \midrule
    \footnotesize
    Parts necessary to maintain a common standard in the ecosystem and at the market
    \\ 
    \footnotesize
    Parts that may allow for a first-mover advantage, if in the interest of the firm
    \\ 
    \footnotesize
    Parts that may force a competitor to adapt
    \\ 
    \footnotesize
    Parts required to maintain or reach a certain level in the ecosystem governance hierarchy
    \\ 
    \midrule
    \footnotesize
\textbf{Relation \& Reputation Drivers:}
    \\ 
    \midrule
    \footnotesize
    Parts that may add to the firm’s reputation as a competitive edge
    \\ 
    \footnotesize
    Parts necessary to maintain relationships with ecosystem participants or external partners
    \\ 
    \footnotesize
    Parts needed to maintain an open attitude internally of firm
    \\ 
    \footnotesize
    Parts necessary to maintain interest among ecosystem participants and attract others
    \\ 
    \footnotesize
    Parts necessary to maintain competitive edge to other OSS ecosystems
    \\ 
    \footnotesize
    Parts necessary to maintain legitimacy and goodwill among ecosystem participants
    \\ 
    \footnotesize
    Parts requested by ecosystem participants and customers
    \\ 
    \footnotesize
    Parts that may help to identify potential employees
    \\ 
    \midrule
    \footnotesize
\textbf{Differential Drivers:}
    \\ 
    \midrule
    \footnotesize
    Parts that may enable internally differential parts
    \\ 
    \footnotesize
    Parts that are non-differentiating for possible competitors in the ecosystem
    \\ 
    \footnotesize
    Parts regarded as commodity
    \\ \bottomrule
\end{tabular}
\vspace{-0.5cm}
\end{table}

\vspace{-0.3cm}
\section{Discussion and Conclusions}
\vspace{-0.3cm}
Target audience for the framework are firms engaged in OSS ecosystems. We believe that the interaction between the firm and the framework should be managed in a workshop format. Further, as in the traditional roadmapping process~\cite{Komssi2015}, we believe that the participants should be cross-functional and include those concerned with the use of OSS in the firm, e.g., legal, management, marketing, product managers, project managers, community managers, and developers. In the workshop, each level of the framework should be addressed and discussed to create a unified view of the current state of practice and how it can be optimized in order for the different levels of the framework to align. 

By identifying its contribution drivers, firms may understand what the alternative cost is to not contribute back to an OSS project and its ecosystem. By aligning this to their internal product planning and business requirement, we believe that they can motivate what should be contributed and not. 
For some firms this may be part of an improvement and maturity process in which the firm starts to understand how they should act in order to influence and draw value from the OSS project and its ecosystem. With time firms may realize how they can make use of OSS projects and their ecosystems on a general level, e.g., how to adapt business models, but also to adopt new ones. 

Pending on the ecosystem’s stakeholder population, there may be multiple agendas present, all of which may not align. 
The agendas may reveal potential competitors and partners, of which some stakeholders may hold both roles as typical in co-opetition. All these factors impact how an OSS project and its ecosystem should be used and engaged~\cite{munir2015open}. This highlights the importance for stakeholder identification and analysis processes to be in place in order to provide necessary input when working with the framework and the improvement process. By using the same framework in the analysis to profile other stakeholders, a firm may benchmark and learn more about how they can adjust their requirements scoping, e.g., in order to gain better influence in the ecosystem.

In this paper we have created an initial version of our framework based on one design cycle. The initial validation highlights the importance of stakeholder identification and analysis, the need for suggested alignment as well as for context specific contribution strategies in general. 

In future work, we plan to reiterate the framework using further expert validation, and also develop an initial set of workshop guidelines for how the framework may be used in an interactive manner. After stabilization is reached in the structure of the framework, it will be piloted in a workshop format with firms engaged in OSS ecosystems. Consideration will be taken to background of experts and firms to strengthen external validity, e.g., in regards to size of development organization and usage of OSS in relation to the business model of firm, but also type of OSS project and ecosystem population. The long term goal is to create a strategic support for contribution management based on the proposed framework in this research preview. The strategic support should allow for tailored contribution strategies to be created, communicated and followed-up through the development organization. The support should further take input from continuous stakeholder identification and analysis of the concerned ecosystems. The work aims to help firms engaged in OSS ecosystems to gain the influence needed in order to impose their agenda, align their internal RE with the ecosystem's RE and to maximize their ROI.

\vspace{-0.3cm}
\bibliographystyle{unsrt}
\bibliography{bibliography_master}

\end{document}